\documentclass[]{elsart}
\usepackage{graphicx}
\usepackage{txfonts}
\usepackage{amsbsy}
\renewcommand{\matrix}[1]{\mathbf{#1}}
\renewcommand{\vec}[1]{\mathbf{#1}}
\newcommand{\tildevec}[1]{\tilde{\mathbf{#1}}}
\newcommand{\hatvec}[1]{\hat{\mathbf{#1}}}
\begin{document}
%
\def\aj{AJ }%
\def\araa{ARA\&A }%
\def\apj{ApJ }%
\def\apjl{ApJ }%
\def\apjs{ApJS }%
\def\ao{Appl.~Opt. }%
\def\apss{Ap\&SS }%
\def\aap{A\&A }%
\def\aapr{A\&A~Rev. }%
\def\aaps{A\&AS }%
\def\azh{AZh }%
\def\baas{BAAS }%
\def\jrasc{JRASC }%
\def\memras{MmRAS }%
\def\mnras{MNRAS }%
\def\pra{Phys.~Rev.~A }%
\def\prb{Phys.~Rev.~B }%
\def\prc{Phys.~Rev.~C }%
\def\prd{Phys.~Rev.~D }%
\def\pre{Phys.~Rev.~E }%
\def\prl{Phys.~Rev.~Lett. }%
\def\pasp{PASP }%
\def\pasj{PASJ }%
\def\qjras{QJRAS }%
\def\skytel{S\&T }%
\def\solphys{Sol.~Phys. }%
\def\sovast{Soviet~Ast. }%
\def\ssr{Space~Sci.~Rev. }%
\def\zap{ZAp }%
\def\nat{Nature }%
\def\iaucirc{IAU~Circ. }%
\def\aplett{Astrophys.~Lett. }%
\def\apspr{Astrophys.~Space~Phys.~Res. }%
\def\bain{Bull.~Astron.~Inst.~Netherlands }%
\def\fcp{Fund.~Cosmic~Phys. }%
\def\gca{Geochim.~Cosmochim.~Acta }%
\def\grl{Geophys.~Res.~Lett. }%
\def\jcp{J.~Chem.~Phys. }%
\def\jgr{J.~Geophys.~Res. }%
\def\jqsrt{J.~Quant.~Spec.~Radiat.~Transf. }%
\def\memsai{Mem.~Soc.~Astron.~Italiana }%
\def\nphysa{Nucl.~Phys.~A }%
\def\physrep{Phys.~Rep. }%
\def\physscr{Phys.~Scr }%
\def\planss{Planet.~Space~Sci. }%
\def\procspie{Proc.~SPIE }%
\let\astap=\aap
\let\apjlett=\apjl
\let\apjsupp=\apjs
\let\applopt=\ao

\begin{frontmatter}

\title{Absorption and scattering properties of arbitrarily shaped particles in the Rayleigh domain}
\subtitle{\large A rapid computational method and a theoretical foundation for the statistical approach}


\author[Pannekoek]{M. Min,}
\ead{mmin@science.uva.nl} 
\author[Pannekoek]{J.~W. Hovenier,}
\author[Pannekoek]{C. Dominik,}
\author[Pannekoek]{A. de Koter,}
\author[Computational science,Chemical kinetics]{M.~A. Yurkin}
\address[Pannekoek]{Astronomical institute Anton Pannekoek, University of Amsterdam, Kruislaan 403, 1098 SJ, Amsterdam, The Netherlands}
\address[Computational science]{Section Computational Science, University of Amsterdam, Kruislaan 403, 1098 SJ, Amsterdam, The Netherlands}
\address[Chemical kinetics]{Institute of Chemical Kinetics and Combustion, Siberian Branch of the Russian Academy of Sciences, Institutskaya 3, 630090, Novosibirsk, Russia}
\date{Last revision \today}

\begin{abstract}
We provide a theoretical foundation for the statistical approach for computing the absorption properties of particles in the Rayleigh domain.  
We present a general method based on the Discrete Dipole Approximation (DDA) to compute the absorption and scattering properties of particles in the Rayleigh domain. The method allows to separate the geometrical aspects of a particle from its material properties. Doing the computation of the optical properties of a particle once, provides them for any set of refractive indices, wavelengths and orientations. This allows for fast computations of e.g. absorption spectra of arbitrarily shaped particles. Other practical applications of the method are in the interpretation of atmospheric and radar measurements as well as computations of the scattering matrix of small particles as a function of the scattering angle.
In the statistical approach, the optical properties of irregularly shaped particles are represented by the average properties of an ensemble of particles with simple shapes.  We show that the absorption cross section of an ensemble of arbitrarily shaped particles with arbitrary orientations can always be uniquely represented by the average absorption cross section of an ensemble of spheroidal particles with the same composition and fixed orientation. This proves for the first time that the statistical approach is generally viable in the Rayleigh domain.
\end{abstract}

\end{frontmatter}
%

\section{Introduction}
\label{sec:Introduction}

The absorption and scattering properties of small particles are very important in both astronomical and atmospheric remote sensing applications.
The interaction of light with particles much smaller than the wavelength of radiation has been studied first by Lord Rayleigh \cite{Rayleigh1871} who explained from basic physical principles the color and polarization of the light from the sky. When studying the detailed spectral properties of the interaction of particles with light, we have to consider the effects of particle size, shape and composition. In this paper we discuss the effects of shape and composition of homogeneous particles, while taking the particle sizes to be in the Rayleigh domain, i.e. much smaller than the wavelength both inside and outside the particle.

When considering the possibilities for computing the optical (i.e. the absorption and scattering) properties of small particles there are two extreme approaches one can take. The first approach is to assume the particles are homogeneous in composition and spherical in shape, which allows us to perform fast and simple computations of its interaction with light using Mie theory \cite{Mie}. The second extreme is to make a model of the particle in an exact way, and perform numerical calculations to obtain its optical properties. This can be done using, for example, the so-called DDA (Discrete Dipole Approximation; see e.g. \cite{1988ApJ...333..848D}) or the T-matrix method (see e.g. \cite{Mishchenko1996a,Wriedt2002}). The first approach is fast, and provides insight into the physics and effects that play a role in the interaction of light with small particles (see e.g. \cite{vandeHulst}). However, due to the perfect symmetry of homogeneous spheres, resonance effects may occur, for example, at particular values of the refractive index, that are not seen in realistically shaped natural particles. This limits the applicability of this approach. The second approach allows us to reproduce details in the observed properties of irregular particles. The main drawback is, however, that the computational demand of most numerical techniques available to compute the optical properties of realistically shaped particles, is high. 
If we wish to consider a large collection of various particle compositions, sizes or wavelengths, one can resort to a third method, the statistical approach. In the statistical approach one simulates the average optical properties of an ensemble of irregularly shaped particles by the average properties of an ensemble of particles with simple shapes. These simple shapes guarantee computations of the optical properties to be relatively fast. In addition, by choosing a broad distribution of simple shapes, we can get rid of the resonance effects which can occur when using homogeneous spheres. The statistical approach is proven useful for, for example, computing absorption spectra of small forsterite grains \cite{2003A&A...404...35M}, and for calculating the degree of linear polarization for small quartz particles \cite{MinHollow}. Some of the successes and limitations of the statistical approach are discussed in refs.~\cite{2003A&A...404...35M,MinHollow,1984ApOpt..23.1025H,1997JGR...10213543M,2002JQSRT..74..167K,2004JQSRT..85..231K}.

In this paper we employ an analytical method based on the DDA to compute the optical properties of particles in the Rayleigh domain in an efficient way. This method is then used to show that the average absorption cross section of an ensemble of particles with arbitrary shapes and orientations is identical to that of an ensemble of particles with spheroidal shapes in a fixed orientation and with the same composition. We thus provide an analytical basis for the use of the statistical approach in computing the absorption cross sections of particles in the Rayleigh domain.

In Sect.~\ref{sec:The absorption properties} we outline the method which is summarized for practical purposes in Sect.~\ref{sec:summary of the procedure}. In Sect.~\ref{sec:Application to various shapes} a few examples of applications to complex particles shapes are provided. A discussion of the implications for the applicability of the statistical approach is given in Sect.~\ref{sec:discussion}. In this section we will also provide suggestions on how to use the derived shape distribution of spheroids to calculate the optical properties of particles with sizes outside the Rayleigh domain.

\section{The absorption properties of very small particles}
\label{sec:The absorption properties}

In this section we will outline the method to obtain the optical properties of arbitrarily shaped and arbitrarily oriented particles small compared to the wavelength. We do this using a solution of the DDA. The general equations for the optical properties of particles in the Rayleigh domain are given in Sect.~\ref{sec:General equations in the Rayleigh domain}. The method to compute the optical properties for arbitrarily shaped particles is outlined in Sect.~\ref{sec:The Discrete Dipole Approximation (DDA)} and summarized in Sect.~\ref{sec:summary of the procedure}.
In Sect.~\ref{sec:formfactors} we introduce the \emph{distribution of form-factors}. Using this distribution we prove the following fundamental theorem:
\begin{thm}
\label{the:statistical approach}
The absorption cross section of an arbitrarily shaped and arbitrarily oriented, homogeneous particle in the Rayleigh domain, or an ensemble of such particles with various shapes and orientations, equals the average cross section of an ensemble of spheroidal particles in a fixed orientation with the same composition and a shape distribution that is independent of the composition of the particle.
\end{thm}
This theorem proves the validity of the statistical approach for calculations of the absorption cross sections for particles in the Rayleigh domain. It is also an encouraging result for applications of the statistical approach for other purposes.

\subsection{General equations in the Rayleigh domain}
\label{sec:General equations in the Rayleigh domain}

In the Rayleigh domain a particle in a given orientation interacts with incident light as a single dipole with dipole moment $\vec{p}$ given by
\begin{equation}
\vec{p}=\boldsymbol{\alpha} \vec{E}_\mathrm{inc}\,\,.
\end{equation}
In this equation $\boldsymbol{\alpha}$ is the $3\times 3$ polarizability tensor, and $\vec{E}_\mathrm{inc}$ is the incoming electric field. From the polarizability tensor it is possible to obtain all scattering and absorption properties of the particle (see e.g. \cite{BohrenHuffman}). The absorption cross section of such a particle when the incident field is applied along the $x,y$ or $z$ axis of a Cartesian coordinate system (indicated by the symbol $\mu=1,2,3$) is given in ref.~\cite{Mackowski1995}
\begin{equation}
\label{eq:oriented cabs}
C_\mathrm{abs}^{(\mu)}=k~\mathrm{Im}\left(\alpha^{(\mu\mu)}\right),
\end{equation}
and the scattering cross section by
\begin{equation}
\label{eq:oriented csca}
C_\mathrm{sca}^{(\mu)}=\frac{k^4}{6\pi}~\sum_{\nu=1}^3 \left|\alpha^{(\mu\nu)}\right|^2,
\end{equation}
where $\alpha^{(\mu\nu)}$ is the $\mu,\nu$-th component of the polarizability tensor ($\mu,\nu=1,2,3$), and $k=2\pi/\lambda$, with $\lambda$ the wavelength of incident radiation. 
Note that when a particle is in the Rayleigh domain the electric field can be taken constant over the particle volume. This implies that the direction of wave propagation of the incident light is not important for the interaction of the light with the particle, but only the direction of the electric field.
The cross sections averaged over all particle orientations (denoted by $\left<...\right>$) are simply given by the average over the three axes which results in
\begin{equation}
\label{eq:cabs}
\left<C_\mathrm{abs}\right>=\frac{k}{3}~\mathrm{Im}\left(\sum_{\mu=1}^3\alpha^{(\mu\mu)}\right),
\end{equation}
and
\begin{equation}
\label{eq:csca}
\left<C_\mathrm{sca}\right>=\frac{k^4}{18\pi}~\sum_{\mu,\nu=1}^3 \left|\alpha^{(\mu\nu)}\right|^2.
\end{equation}

For only a few particle shapes simple analytical equations are available to compute the polarizability tensor. These are the shapes of homogeneous, layered and hollow spheres, and homogeneous, layered and hollow ellipsoids.
For a homogeneous sphere $\boldsymbol{\alpha}$ is given in ref.~\cite{BohrenHuffman}
\begin{equation}
\label{eq:alpha sphere}
\boldsymbol{\alpha}=3V\frac{m^2-1}{m^2+2}~\matrix{I_\mathrm{3}}\,\,,
\end{equation}
with $m$ the complex refractive index of the particle material, $V$ the material volume of the particle, and $\matrix{I_\mathrm{3}}$ the $3\times 3$ identity matrix.
The polarizability tensor of a homogeneous ellipsoidal particle is given by ref.~\cite{BohrenHuffman}
\begin{equation}
\label{eq:Rayleigh ellips}
\boldsymbol{\alpha}=V\left[\begin{array}{ccc}
\frac{m^2-1}{1+L_1(m^2-1)} & 0 & 0 \\
0 & \frac{m^2-1}{1+L_2(m^2-1)} & 0 \\
0 & 0 & \frac{m^2-1}{1+L_3(m^2-1)} \\
\end{array}\right],
\end{equation}
where the $L_i$ are geometrical form-factors with values between $0$ and $1$ depending on the shape of the ellipsoid.

For an ellipsoid of revolution (a spheroid) with axes $a$ and $b$ we have
\begin{equation}
\label{eq:L values}
L_1=\left\{ \begin{array}{lll}  
\displaystyle{\frac{1-e^2}{e^2}\left(-1+\frac{1}{2e}\ln\left(\frac{1+e}{1-e}\right)\right),} & \quad\displaystyle{e^2=1-\frac{a^2}{b^2},} & \quad\mathrm{Prolates,} \\ &\\
\displaystyle{\frac{1}{e^2}\left(1-\frac{\sqrt{1-e^2}}{e}\arcsin(e)\right),} & \quad\displaystyle{e^2=1-\frac{b^2}{a^2},} & \quad\mathrm{Oblates,} \end{array}\right.
\end{equation}
and $L_2=L_3=(1-L_1)/2$. Here $b$ is the rotation axis, and $a/b$ is the aspect ratio. For prolate spheroids $a/b<1$ and $0<L_1<1/3$, while for oblate spheroids $a/b>1$ and $1/3<L_1<1$. For homogeneous spheres $L_1=L_2=L_3=1/3$. 

The equations for the polarizability of hollow spheres or ellipsoids can be found in ref.~\cite{vandeHulst}. 
For a hollow sphere having a volume fraction $f$ occupied by a central vacuum inclusion, the polarizability tensor is given by
\begin{equation}
\label{eq:Hollow sphere}
\boldsymbol{\alpha}=3V\frac{(m^2-1)(2m^2+1)}{(m^2+2)(2m^2+1)-2(m^2-1)^2f}~\matrix{I_\mathrm{3}}\,\,,
\end{equation}
where $V=\frac{4}{3}\pi r^3(1-f)$ is the material volume of a hollow sphere with outer radius $r$.

For convenience we introduce the dimensionless polarizability per unit material volume of a particle $\overline{\boldsymbol{\alpha}}=\boldsymbol{\alpha}/V$.

\subsection{The Discrete Dipole Approximation (DDA)}
\label{sec:The Discrete Dipole Approximation (DDA)}

In the Discrete Dipole Approximation (DDA) a particle is represented by a collection of interacting dipoles. This idea was proposed by Purcell \& Pennypacker \cite{1973ApJ...186..705P}, who derived the general equations for the interaction of the dipoles with the incident light and with each other. The approach is equivalent to a discretization of the particle volume and the assumption that each volume element interacts with the radiation field as a single dipole $j$ with polarizability $\boldsymbol{\beta}_j$. 
For a theoretical derivation of this approach from the Maxwell equations see e.g. \cite{1992ApJ...394..494L,Lakhtakia1993}.
For inhomogeneous particles the polarizability can vary throughout the particle depending on the material properties of the local volume element. Also the polarizability tensor can be anisotropic. In this paper we consider only homogeneous particles, i.e. $\boldsymbol{\beta}_j=\boldsymbol{\beta}$ for all $j$. In addition we consider only the case when the polarizability is isotropic in which case the polarizability tensor can be written as $\beta\matrix{I_\mathrm{3}}$, with $\beta$ the scalar polarizability. For the DDA to be valid, the volume elements represented by the dipoles have to be in the Rayleigh domain.

To compute the interaction of $N$ dipoles with the incident field and with each other we have to solve a set of $N$ inhomogeneous linear vector equations, namely
\begin{equation}
\label{eq:general DDA 1}
\sum_{j=1}^{N}(\matrix{A_\mathit{ij}}+\beta^{-1}\matrix{I_\mathrm{3}})\vec{P}_j=\vec{E}_{\mathrm{inc},i}\qquad (i=1,...,N).
\end{equation}
Here the polarization vector at the position of dipole $j$, $\vec{P}_j$, is a column vector with $3$ elements. $\vec{E}_{\mathrm{inc},i}$ is a column vector representing the incoming electric field at the position of dipole $i$. The $\matrix{A_\mathit{ij}}$ are ${3\times 3}$ matrices determined by the positions of the dipoles in the particle and thus by the particle geometry. The general equations for the $\matrix{A_\mathit{ij}}$ are given in ref.~\cite{1988ApJ...333..848D}. We assume in this paper that the entire particle is in the Rayleigh domain. This implies that the electric field is constant, so we are in the limit of $k\rightarrow 0$. Then
\begin{equation}
\label{eq:Rayleigh DDA elements}
\matrix{A_\mathit{ij}}=\left\{ \begin{array}{ll}
\displaystyle{\frac{\matrix{I_\mathrm{3}}-3{\hatvec{r}}_{ij}{\hatvec{r}}_{ij}}{4\pi r_{ij}^3}
,} & \quad\displaystyle{i\neq j,} \\ 
&\\
\displaystyle{0,} & \quad\displaystyle{i=j.} \end{array}\right.
\end{equation}
Here ${\hatvec{r}}_{ij}$ is the unit vector pointing from dipole $i$ to dipole $j$, and $r_{ij}$ is the distance between these two dipoles. The $3\times 3$ matrix ${\hatvec{r}}_{ij}~{\hatvec{r}}_{ij}$ is defined as the product of $\hatvec{r}_{ij}$ as a column vector and $\hatvec{r}_{ij}$ as a row vector, i.e.
\begin{equation}
{\hatvec{r}}_{ij}~{\hatvec{r}}_{ij}=\left[\begin{array}{c@{\qquad}c@{\qquad}c}
\hat{r}_{ij,x}^2 			& \hat{r}_{ij,x}~\hat{r}_{ij,y}	& \hat{r}_{ij,x}~\hat{r}_{ij,z}	\\
\hat{r}_{ij,x}~\hat{r}_{ij,y}	& \hat{r}_{ij,y}^2 			& \hat{r}_{ij,y}~\hat{r}_{ij,z}	\\
\hat{r}_{ij,x}~\hat{r}_{ij,z}	& \hat{r}_{ij,y}~\hat{r}_{ij,z}	& \hat{r}_{ij,z}^2 			\\
\end{array}\right],
\end{equation}
where $\hat{r}_{ij,x},\hat{r}_{ij,y}$ and $\hat{r}_{ij,z}$ are the $x,y$ and $z$ components of the unit vector ${\hatvec{r}}_{ij}$, respectively.

Following \cite{1988ApJ...333..848D} we define the $3N$ dimensional column vectors ${\tildevec{E}}_\mathrm{inc}=(\vec{E}_{\mathrm{inc},1},$ $\vec{E}_{\mathrm{inc},2},$ $...,$ $\vec{E}_{\mathrm{inc},N})$ and ${{\tildevec{P}}=(\vec{P}_1, \vec{P}_2, ..., \vec{P}_N)}$ and the ${3N\times 3N}$ matrix $\matrix{\tilde{A}}$ such that $\matrix{\tilde{A}_\mathit{\mathrm{3}(i-\mathrm{1})+\mu,\mathrm{3}(j-\mathrm{1})+\nu}}=(\matrix{A_\mathit{ij}})_{\mu\nu}$, $\{i,j=1...N; \mu,\nu=1,2,3\}$. In this way Eq.~(\ref{eq:general DDA 1}) can be written as a single matrix equation
\begin{equation}
\label{eq:general DDA}
(\matrix{\tilde{A}}+\beta^{-1}\matrix{I_\mathit{\mathrm{3}N}})\tildevec{P}=\tildevec{E}_{\mathrm{inc}}\,\,.
\end{equation}
Note that in this equation the matrix $\matrix{\tilde{A}}$ only depends on the particle geometry, while the matrix $\beta^{-1}\matrix{I_\mathit{\mathrm{3}N}}$ only depends on the dielectric properties of the particle.

Eq.~(\ref{eq:general DDA}) can be solved for $\tildevec{P}$ using, for example, a conjugate gradient method. In the literature considerable effort has been reported to increase the efficiency of solution methods for Eq.~(\ref{eq:general DDA}) (see e.g. \cite{Goodman1991,Hoekstra}).
In contrast to the general equations for $\matrix{\tilde{A}}$, in the Rayleigh domain this matrix is real and symmetric. This implies that the eigenvalues of $\matrix{\tilde{A}}$ are real, and that the eigenvectors form an orthonormal basis. As we will detail below, once the eigenvalues and eigenvectors of $\matrix{\tilde{A}}$ have been determined, the solution of Eq.~(\ref{eq:general DDA}) can be obtained for arbitrary values of $\beta$, and thus for any particle material. This method is similar to that presented by Markel~et~al. \cite{1991PhRvB..43.8183M}.

Using the eigenvalues and eigenvectors, the matrix $\matrix{\tilde{A}}$ can be diagonalized by writing
\begin{equation}
\matrix{\tilde{A}}=\matrix{UDU^\mathit{T}},
\end{equation}
where a superscript $T$ is used to denote the transpose of a matrix. The columns of $\matrix{U}$ are the $3N$ real valued eigenvectors of $\matrix{\tilde{A}}$, and the elements of the diagonal matrix $\matrix{D}$ are the corresponding eigenvalues of $\matrix{\tilde{A}}$.
Since the eigenvectors of $\matrix{\tilde{A}}$ are orthonormal the transpose of $\matrix{U}$ equals its inverse, i.e. $\matrix{U^\mathrm{T}}=\matrix{U^\mathrm{-1}}$. The eigenvalues and eigenvectors of a real symmetric matrix can be found relatively easy using numerical techniques.

The matrix on the left hand side of Eq.~(\ref{eq:general DDA}) can be rewritten after diagonalization since
\begin{equation}
(\matrix{\tilde{A}}+\beta^{-1}\matrix{I_\mathit{\mathrm{3}N}})=(\matrix{UDU^\mathit{T}}+\beta^{-1}\matrix{I_\mathit{\mathrm{3}N}})=\matrix{U}(\matrix{D}+\beta^{-1}\matrix{I_\mathit{\mathrm{3}N}})\matrix{U^\mathit{T}}.
\end{equation}
Using this, the solution of Eq.~(\ref{eq:general DDA}) can be written as
\begin{equation}
\label{eq:Polarize}
\tildevec{P}=\matrix{U}(\matrix{D}+\beta^{-1}\matrix{I_\mathit{\mathrm{3}N}})^{-1}\matrix{U^\mathit{T}}~\tildevec{E}_\mathrm{inc}\,\,.
\end{equation}
Since $(\matrix{D}+\beta^{-1}\matrix{I_\mathit{\mathrm{3}N}})$ is a diagonal matrix, its inverse is readily computed. Once $\matrix{U}$ and $\matrix{D}$ have been calculated, the solution can be easily computed for every value of $\beta$. Since $\matrix{U}$ and $\matrix{D}$ only depend on the geometry of the particle, we only have to calculate them once in order to find the solution for arbitrary values of $\beta$, and thus for any particle material.

In the Rayleigh domain, the incident electric field constant throughout the particle. This means that, at each dipole $j$, $\vec{E}_{\mathrm{inc},j}=\vec{E}_\mathrm{inc}$. The total dipole moment of the particle then is
\begin{equation}
\vec{p}=\sum_{j=1}^{N}\vec{P}_j=\boldsymbol{\alpha} \vec{E}_\mathrm{inc}\,\,.
\end{equation}
Combining this equation with Eq.~(\ref{eq:Polarize}) gives us the polarizability tensor $\boldsymbol{\alpha}$ with elements
\begin{equation}
\label{eq:final polarizability}
\alpha^{(\mu\nu)}=\sum_{j=1}^{3N} \frac{w_j^{(\mu\nu)}}{\lambda_j+\beta^{-1}}\,,
\end{equation}
where the $\lambda_j$ are the $3N$ eigenvalues of $\matrix{\tilde{A}}$, and the $w_j^{(\mu\nu)}$ are given by
\begin{equation}
\label{eq:weights}
w_j^{(\mu\nu)}=\sum_{i,k=1}^N \matrix{U_\mathit{j,\mathrm{3}(i-\mathrm{1})+\mu}}\matrix{U_\mathit{j,\mathrm{3}(k-\mathrm{1})+\nu}}\,\,\,.
\end{equation}
Note that since the matrix $\matrix{U}$ is real valued, the $w_j^{(\mu\mu)}$ are real valued and positive since
\begin{equation}
\label{eq:wj mumu}
w_j^{(\mu\mu)}=\left(\sum_{i=1}^N \matrix{U_\mathit{j,\mathrm{3}(i-\mathrm{1})+\mu}}\right)^2.
\end{equation}

Using Eqs.~(\ref{eq:oriented cabs}) and (\ref{eq:final polarizability}), the absorption cross section when the field is applied along the $\mu$ axis becomes
\begin{equation}
\label{eq:final Cabs}
C_\mathrm{abs}^{(\mu)}=k~\mathrm{Im}\left(\sum_{j=1}^{3N} \frac{w_j^{(\mu\mu)}}{\lambda_j+\beta^{-1}}\right).
\end{equation}
Also all the scattering properties, i.e. the scattering cross section, matrix, and depolarization factors, of the particle can be obtained from the polarizability tensor. The scattering cross section when the field is applied along the $\mu$ axis can be computed using Eqs.~(\ref{eq:oriented csca}) and (\ref{eq:final polarizability}) and becomes
\begin{equation}
\label{eq:final Csca}
C_\mathrm{sca}^{(\mu)}=\frac{k^4}{6\pi}~\sum_{\nu=1}^3\left|\sum_{j=1}^{3N}\frac{w_j^{(\mu\nu)}}{\lambda_j+\beta^{-1}}\right|^2.
\end{equation}
The orientation averaged cross sections can also be obtained from Eq.~\ref{eq:final polarizability} using Eqs.~(\ref{eq:cabs}) and (\ref{eq:csca}). In the following we focus on the absorption cross section.

\subsection{The distribution of form-factors}
\label{sec:formfactors}

In the previous sections we have shown how to separate the geometrical and dielectrical properties when computing the absorption properties of arbitrarily shaped particles in the Rayleigh domain. In this section we show how the particle geometry can be uniquely described by a distribution of \emph{form-factors}. Also, we will show how this distribution can be used to construct a shape distribution of spheroidal particles in a fixed orientation and with the same composition which has identical absorption properties.

We take the polarizability of a single volume element to be the Clausius-Mossotti polarizability of a cubic volume element with size $d$ \cite{1973ApJ...186..705P}, i.e.
\begin{equation}
\beta=3d^3~\frac{m^2-1}{m^2+2}\,.
\end{equation}
When the size of the volume elements is sufficiently small, this is the polarizability that follows directly from the Maxwell equations (see e.g. \cite{1992ApJ...394..494L,Lakhtakia1993}).

The equation for the absorption cross section (Eq.~\ref{eq:final Cabs}) then reduces to
\begin{equation}
\label{eq:final Cabs2}
C_\mathrm{abs}^{(\mu)}=\sum_{j=1}^{3N} \frac{w_j^{(\mu\mu)}}{N}~\left[kV~\mathrm{Im}\left(\frac{m^2-1}{1+L_j~(m^2-1)}\right)\right],
\end{equation}
with $V=Nd^3$ the total material volume of the particle and the form-factor $L_j$ defined by
\begin{equation}
\label{eq:form-factor}
L_j=\frac{1}{3}+\lambda_j d^3.
\end{equation}

Comparing Eq.~(\ref{eq:final Cabs2}) with Eq.~(\ref{eq:Rayleigh ellips}) we see that Eq.~(\ref{eq:final Cabs2}) is equal to the equation for the average absorption cross section of an ensemble of oriented ellipsoids with various shapes.
We have thus proven theorem \ref{the:statistical approach}.

The result given above implies that we can represent the absorption properties of an arbitrarily shaped and arbitrarily oriented particle uniquely by a distribution of the form-factors, $\mathcal{P}(L)$. But in order to represent an ellipsoid with a given aspect ratio, the $L_j$ have to be between $0$ and $1$. This implies that the eigenvalues, $\lambda_j$, have to be between $-1/(3d^3)$ and $2/(3d^3)$ for arbitrary shapes. For all particle shapes that we considered this was the case.
From the $w_j^{(\mu\mu)}$ (given by Eq.~\ref{eq:wj mumu}) and the $L_j$ (given by Eq.~\ref{eq:form-factor}) it is straightforward to construct the distribution of form-factors.

In practice we often meet ensembles of particles with different shapes and orientations. In the Rayleigh domain, averaging over particle orientations can be done by averaging over three perpendicular orientations. In view of Eq.~(\ref{eq:final Cabs2}) it is clear that the shape and orientation averaged polarizability (and thus the average absorption cross section) of such ensembles can be obtained from the ensemble averaged form-factor distribution. This allows us to construct average form-factor distributions of different classes of particle shapes, resulting in a unique distribution of spheroidal particles to represent the absorption properties of each class.

\subsection{Summary of the procedure}
\label{sec:summary of the procedure}

The method described above provides a practical way of computing the optical properties of arbitrarily shaped and arbitrarily oriented particles in the Rayleigh domain. It separates the geometrical and material properties of the particle. This implies that for a given particle geometry the computation has to be done only once to obtain the absorption and scattering properties for arbitrary values of the refractive index and wavelength.

The procedure to obtain the optical properties of an arbitrarily shaped and arbitrarily oriented particle in the Rayleigh domain is outlined as follows.
\begin{enumerate}
\item Construct the matrix $\matrix{\tilde{A}}$ using Eq.~(\ref{eq:Rayleigh DDA elements}) and $\matrix{\tilde{A}_\mathit{\mathrm{3}(i-\mathrm{1})+\mu,\mathrm{3}(j-\mathrm{1})+\nu}}=(\matrix{A_\mathit{ij}})_{\mu\nu}$.
\item Compute the eigenvalues, $\lambda_j$, and the matrix of eigenvectors, $\matrix{U}$, of the real symmetric matrix $\matrix{\tilde{A}}$. 
\item Compute the $w_j^{(\mu\nu)}$ using Eq.~(\ref{eq:weights}).
\item Compute the polarizability tensor $\boldsymbol{\alpha}$ from Eq.~(\ref{eq:final polarizability}).
\item From the polarizability tensor the absorption and scattering cross sections for the particle in a single orientation are obtained from Eqs.~(\ref{eq:oriented cabs}) and (\ref{eq:oriented csca}). Also, the scattering matrix can be obtained from the polarizability tensor (see \cite{BohrenHuffman})
\item The absorption and scattering cross sections for an ensemble of particles with random orientations can be obtained from Eqs.~(\ref{eq:cabs}) and (\ref{eq:csca}). The scattering matrix for an ensemble of randomly oriented particles is given in ref.~\cite{BohrenHuffman}.
\end{enumerate}
The computationally demanding steps (1), (2) and (3) have to be done only once for a given particle geometry and structure. Steps (4), (5) and (6) then provide the optical properties for arbitrary orientation, wavelength, and particle material.

The distribution of form-factors is computed from the $\lambda_j$  and the $w_j^{(\mu\mu)}$ by using Eq.~(\ref{eq:form-factor}). The corresponding distribution of spheroids is obtained from the form-factor distribution using Eq.~(\ref{eq:L values}). In Sect.~\ref{sec:larger particles} we outline how the distribution of spheroidal particles that is obtained might be used to compute the optical properties of particles outside the Rayleigh domain.

\section{Application to various shapes}
\label{sec:Application to various shapes}

\subsection{Homogeneous spheres and spheroids}

Since the average polarizability of an ensemble of homogeneous spheroids in random orientation consists of the sum of two polarizabilities with different values of $L$ (see Eqs.~\ref{eq:Rayleigh ellips} and \ref{eq:L values}), the ensemble averaged form-factor distribution displays two distinct peaks, one at the value of $L$ determined by Eq.~(\ref{eq:L values}) and one at $(1-L)/2$ (with double intensity). 

\begin{figure}[!t]
\resizebox{\hsize}{!}{\includegraphics{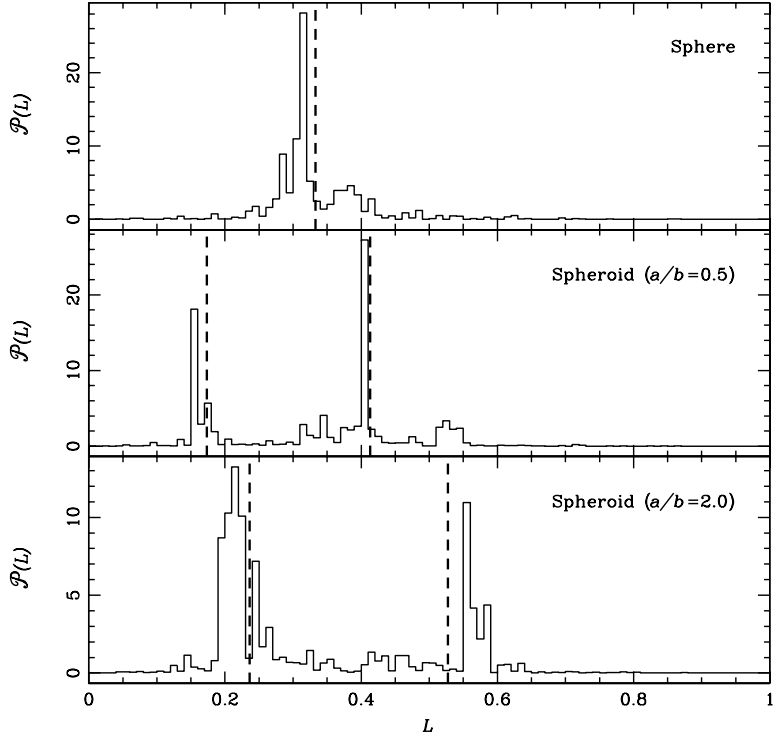}}
\caption{The form-factor distributions as calculated for an array of 3000 dipoles representing a sphere, a prolate spheroid with $a/b=0.5$ and an oblate spheroid with $a/b=2.0$ averaged over all particle orientations. The expected positions of the peaks in the distribution are indicated by the dashed lines.}
\label{fig:ellips distr}
\end{figure}

\begin{figure}[!t]
\resizebox{\hsize}{!}{\includegraphics{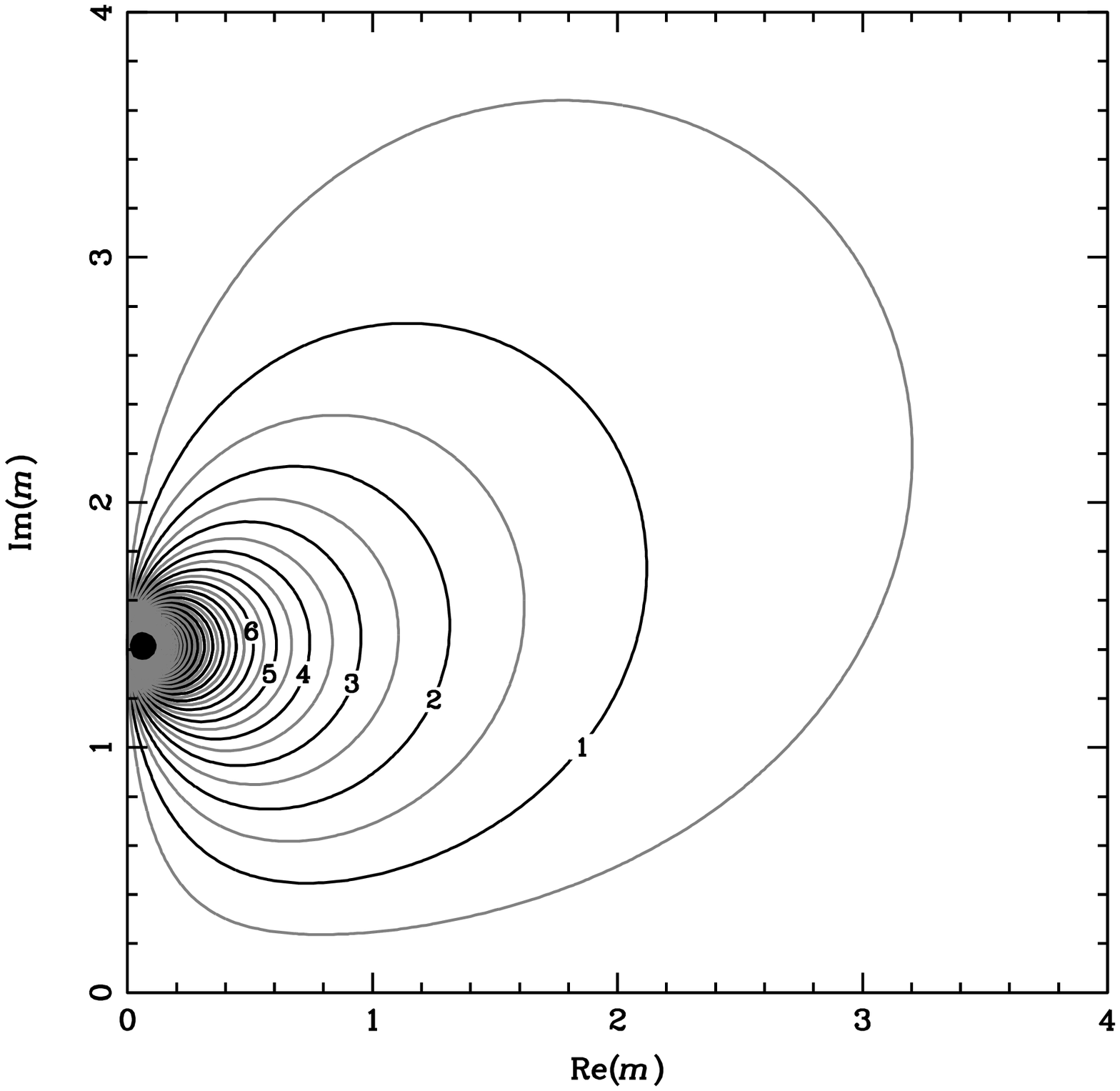}\includegraphics{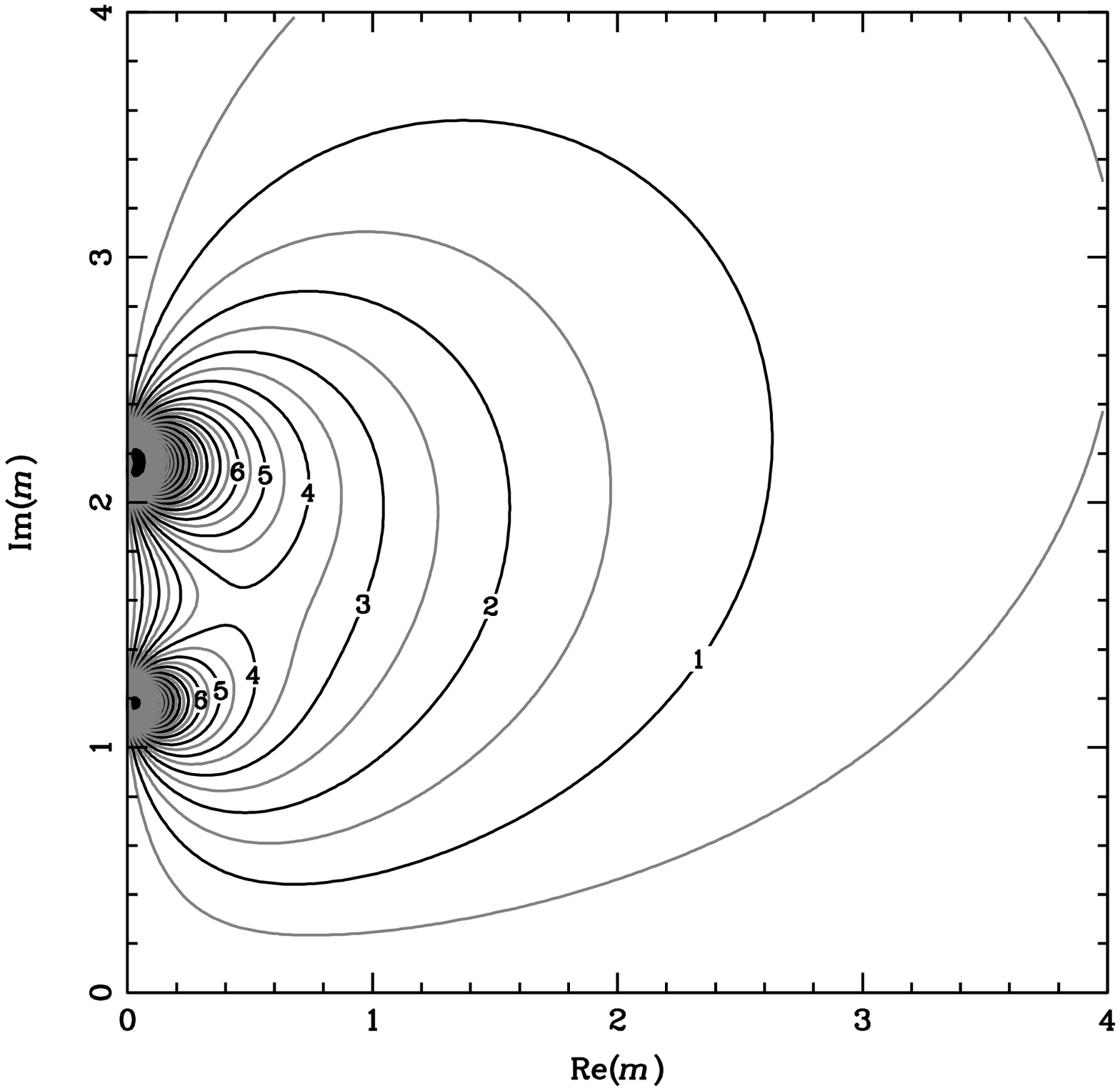}}
\caption{Contour plots of the imaginary part of the polarizability per unit volume as functions of the real and imaginary part of the refractive index. The left panel is for a single homogeneous sphere and the right panel for an ensemble of randomly oriented prolate homogeneous spheroids with aspect ratios $a/b=0.5$.}
\label{fig:sphere spheroid alpha}
\end{figure}

By way of example we computed the form-factor distributions of a homogeneous sphere, a prolate spheroid with aspect ratio $a/b=0.5$ and an oblate spheroid with $a/b=2.0$, averaged over all particle orientations using the DDA method (see Fig.~\ref{fig:ellips distr}). The material volume of each particle is discretized by using $3000$ dipoles. We also indicate in this figure the positions of the peaks as determined by Eq.~(\ref{eq:L values}). We see that, in general, the positions of the peaks in the calculated distributions match the expected values fairly well. The differences are mainly caused by the fact that by using a discrete array of cubic dipoles it is not possible to represent the shape of a spheroid exactly.

The refractive index of natural materials is a function of wavelength. The refractive index of most materials exhibits resonances in the infrared part of the spectrum. These resonances lead to corresponding resonances in the absorption cross section which can be detected, for example, in thermal emission spectra. The wavelength positions of the features are sensitive to the particle shape and can be used to determine the shape and composition of the emitting grains. Therefore, it is important to study the absorption cross section as a function of the refractive index and wavelength.

In the left panel of Fig.~\ref{fig:sphere spheroid alpha} we show $C_\mathrm{abs}/(kV)$ or, in other words, the imaginary part of the polarizability per unit volume, of a single homogeneous sphere as a function of the real and imaginary part of the refractive index as computed from Eq.~(\ref{eq:alpha sphere}). Note the sharp increase of the imaginary part of the polarizability when $m\rightarrow i\sqrt{2}$ typical for homogeneous spherical particles (cf. Eq.~\ref{eq:alpha sphere}). When considering the absorption cross section as a function of wavelength near a resonance, this leads to a sharp peak in the absorption spectrum. The imaginary part of the polarizability per unit volume of an ensemble of randomly oriented prolate spheroids with aspect ratio $a/b=0.5$ is shown in the right panel of Fig.~\ref{fig:sphere spheroid alpha}. In this contour plot we see two sharp maxima, caused by the two values of $L$ determining the polarizability. In general, we expect particles with a few isolated form-factors to show strong resonances for each form-factor. As will be shown below, a full distribution leads to a smoother absorption spectrum.

\subsection{Gaussian random sphere}

In order to model the optical properties of irregularly shaped particles we have to employ a model for the shape of the particles. A successful model that is frequently used in light scattering theory is that of a Gaussian random sphere \cite{1996JQSRT..55..577M}. The basis of this shape model is a homogeneous sphere of which the surface is distorted according to a Gaussian random distribution. The distortion of the surface is parameterized by two shape variables, the standard deviation, $\sigma$, of the distance to the center, and the average correlation angle, $\Gamma$. The value of $\Gamma$ determines the number of hills and valleys on the surface within a solid angle, while $\sigma$ determines the height of these hills and valleys. For details see ref.~\cite{1996JQSRT..55..577M}. Using different shape parameters we can create particles with varying degrees of irregularity. Volten~et~al. \cite{2001JGR...10617375V} successfully used Gaussian random spheres to compute the scattering matrices of mineral particles as functions of the scattering angle. They found that in order to reproduce the experimentally determined scattering behavior they had to employ very large values of $\sigma$.

\begin{figure*}[!t]
\resizebox{\hsize}{!}{\hspace{5cm}\includegraphics{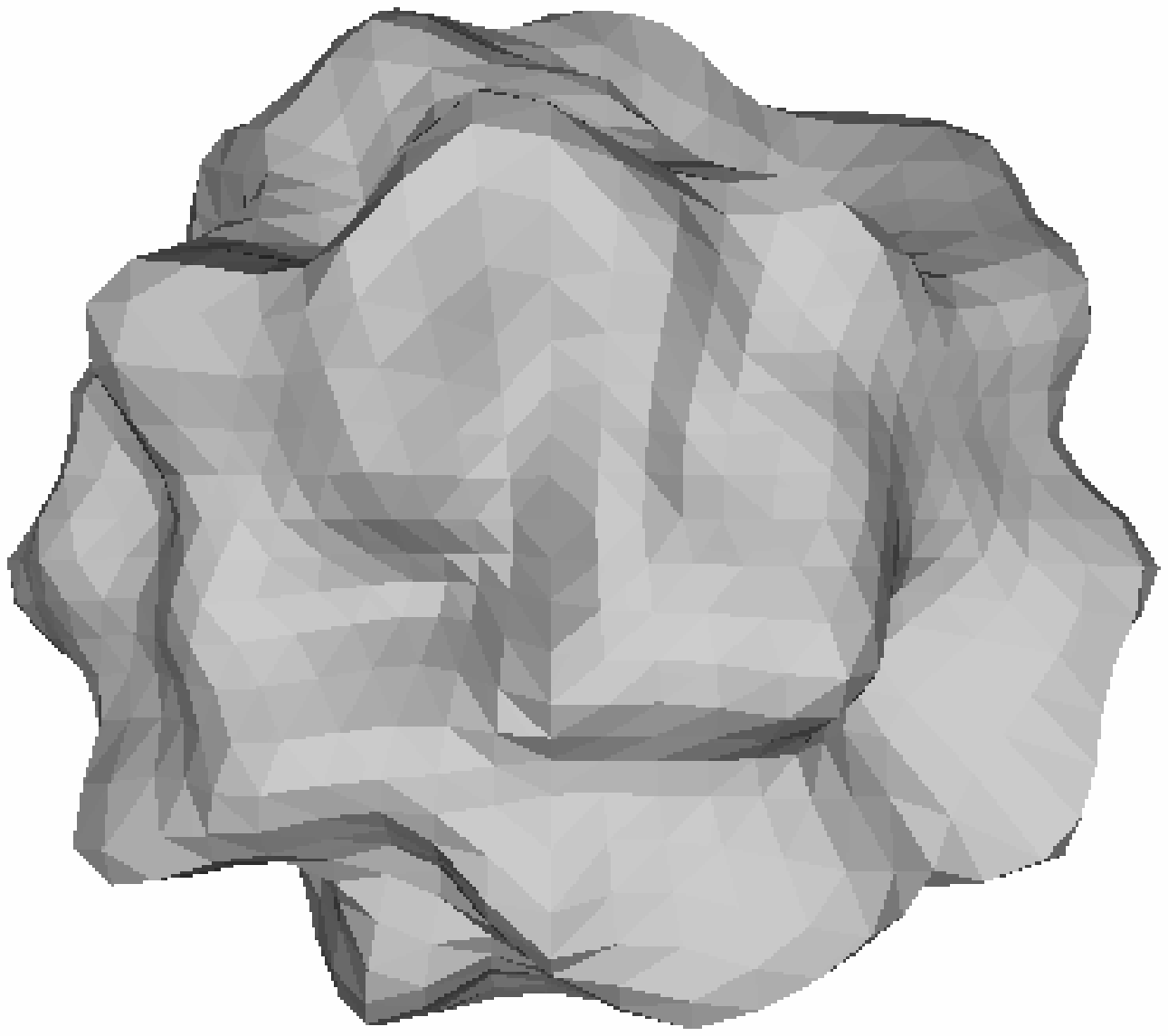}\includegraphics{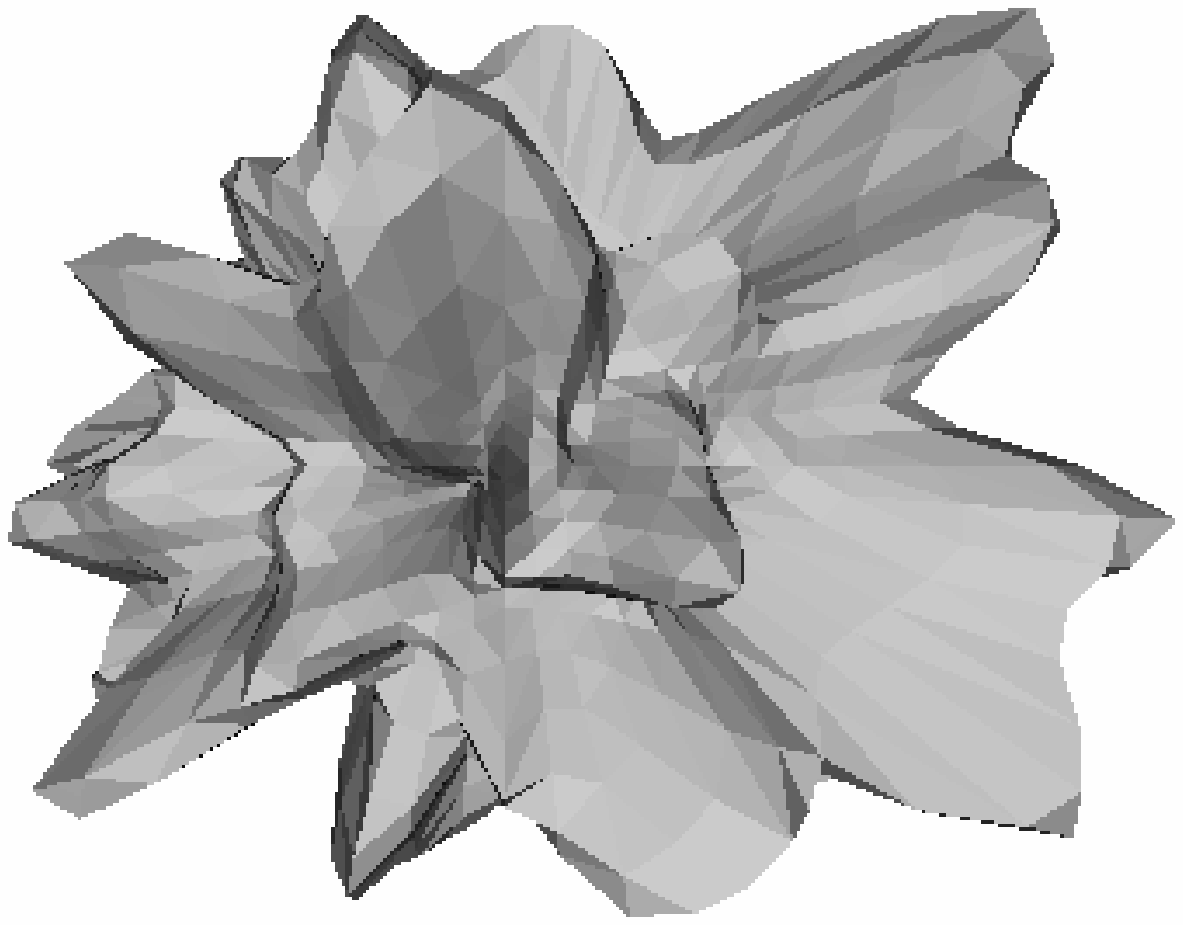}\includegraphics{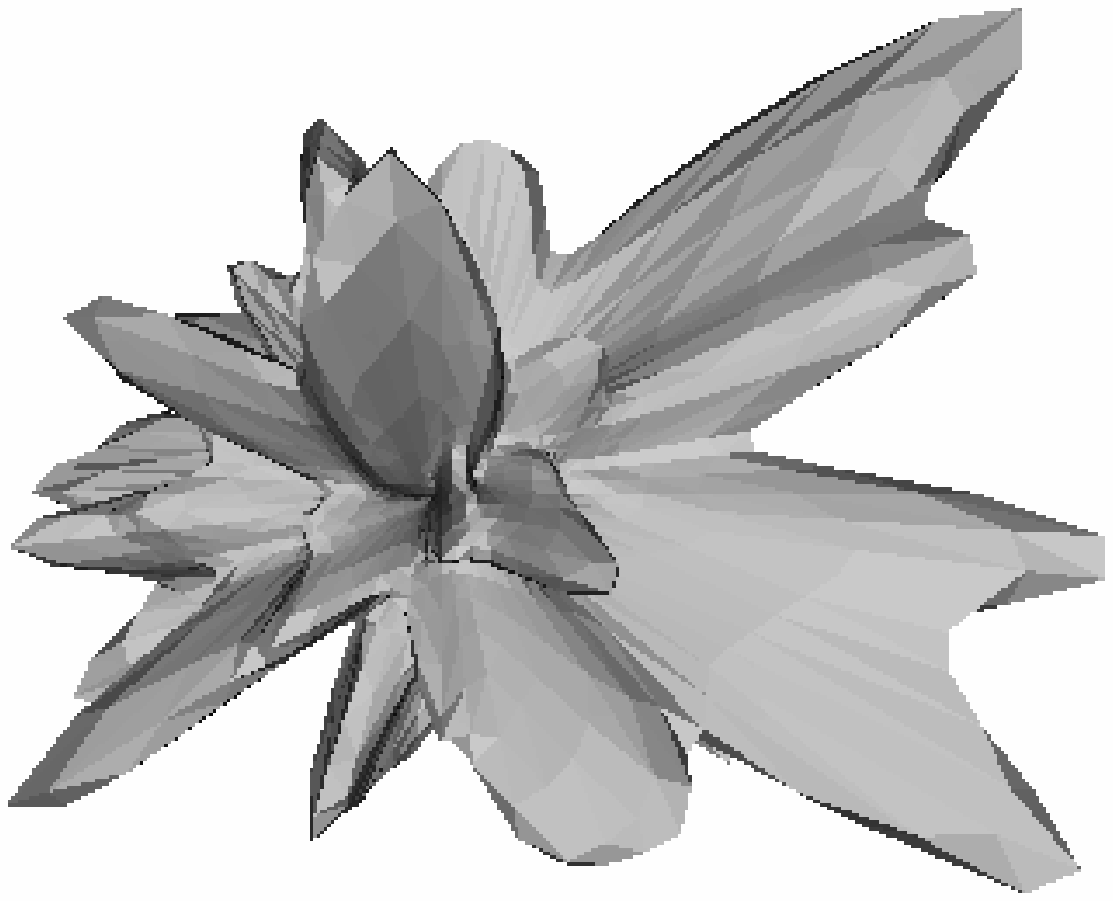}\includegraphics{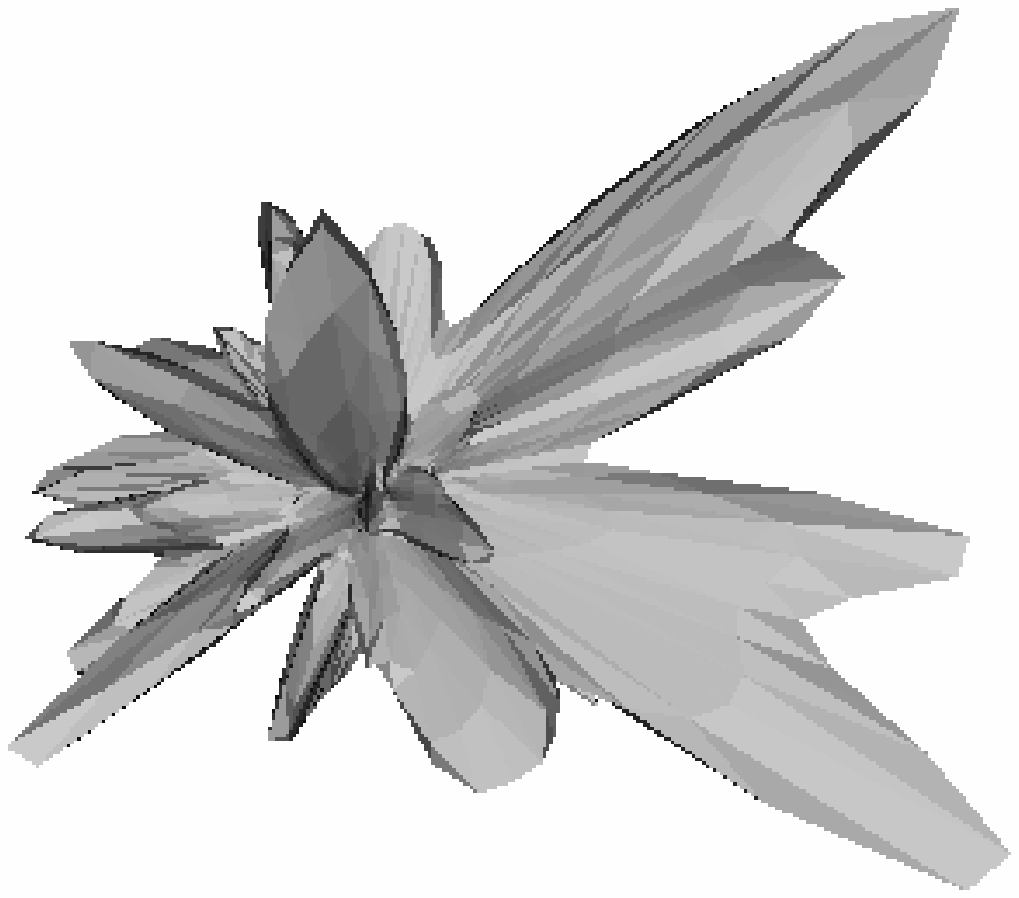}\hspace{1cm}}
\resizebox{\hsize}{!}{\includegraphics{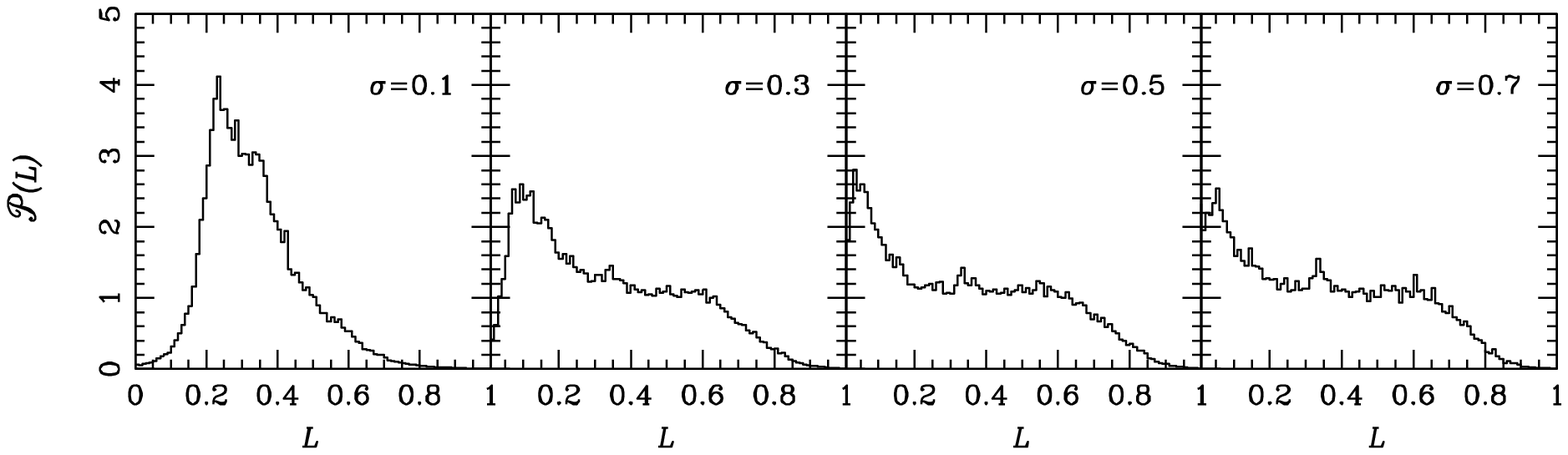}}
\resizebox{\hsize}{!}{\includegraphics{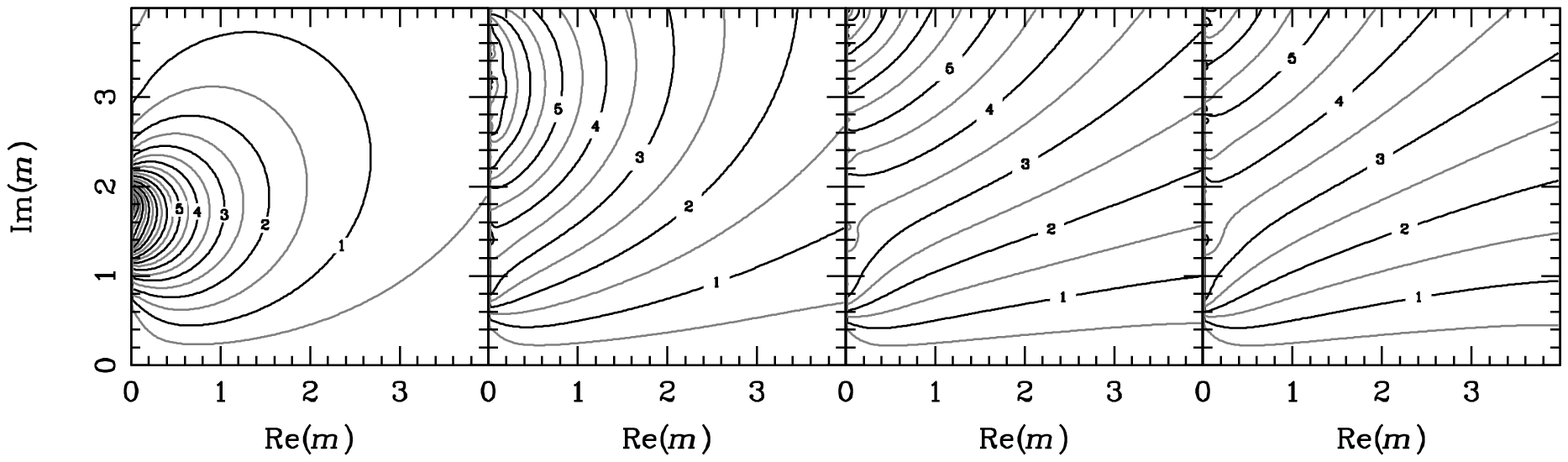}}
\caption{The orientation and ensemble averaged form-factor distribution (top panels) and the imaginary part of the average polarizability per unit volume as a function of the real and imaginary part of the refractive index (bottom panels) of the ensembles of randomly oriented Gaussian random spheres. Above the figures we show typical examples of particles from the various ensembles. From left to right the relative standard deviation of the distance to the center is $0.1$, $0.3$, $0.5$ and $0.7$, respectively. The correlation angle $\Gamma=10^\circ$, and is the same for all four shapes. For each shape we considered an ensemble of $10$ different particles with the same parameters but different seeds of the random number generator. We average over these different particles in order to reduce the effects of random variations of the particle shapes on the final results.}
\label{fig:gaussian spheres}
\end{figure*}

In the top panel of Fig.~\ref{fig:gaussian spheres} we show the form-factor distribution of various ensembles of randomly oriented Gaussian random spheres with different values of $\sigma$. In all cases we fix the value of $\Gamma$ to $10^\circ$. For each value of $\sigma$ we construct an ensemble of 10 different Gaussian random spheres, each with the same value for the shape parameter but different seeds of the random number generator. Above the two panels of Fig.~\ref{fig:gaussian spheres} we show some typical examples from the ensembles. The volume of each particle is discretized using $3000$ dipoles in order to construct the orientation and ensemble averaged form-factor distribution. The form-factor distribution of each ensemble is used to compute its average polarizability per unit volume as a function of the refractive index from Eq.~(\ref{eq:final Cabs2}). The imaginary part of the average polarizability per unit volume is shown by contours in the bottom panel of Fig.~\ref{fig:gaussian spheres}.

Fig.~\ref{fig:gaussian spheres} shows that for small deformations of the surface of the sphere (most left panel), the form-factor distribution is dominated by one peak at $L\approx 1/3$ and the absorption properties are rather similar to those of spherical particles (see the left panel of Fig.~\ref{fig:sphere spheroid alpha}). For larger values of $\sigma$ the distribution of form-factors gets broader. If $\sigma=0.5$ or $0.7$ we see that for a fixed value of $\mathrm{Re}(m)$ the polarizability is a smooth, continuously increasing function of $\mathrm{Im}(m)$, as expected in the absence of resonance effects caused by the geometry of the particle.

In ref.~\cite{Battaglia} a method is proposed to compute the scattering and absorption properties of Gaussian random spheres in the Rayleigh domain. However, the method used in that paper is an approximation using only a single ellipsoid as a best fit to the particle shape. From the discussion given above it is clear that this can provide only a very rough approximation to the real absorption properties of Gaussian random spheres.

\subsection{Aggregates with various fractal dimensions}

In environments where dust particles grow by aggregation of smaller constituents, the resulting grains might be modeled as fractal aggregates. The fractal dimensions of these aggregates are a measure for the compactness or fluffiness of the particle and depend on the conditions in the environment. A large part of the interplanetary dust particles collected in the Earth's atmosphere are aggregates of small particles. Therefore, the optical properties of particle aggregates are very important in astrophysics and planetary physics. Fogel \& Leung \cite{1998ApJ...501..175F} computed emission and extinction spectra of fractal aggregates with various fractal dimensions. They found that the mass absorption and extinction coefficients of fractal aggregates are on average higher than those of volume equivalent spheres. Mackowski \cite{Mackowski1995} developed a method to obtain the optical properties of sphere clusters in the Rayleigh domain. 
For computations of various values of the refractive index, which are needed, for example, for computing absorption spectra, the method we employ is much faster and, in addition, it is easier to implement.

We construct aggregates with various fractal dimensions using a sequential tunable particle-cluster aggregation method developed by Filippov~et~al. \cite{Filippov}. A fractal aggregate composed of homogeneous spheres obeys the so-called scaling law \cite{Filippov}
\begin{equation}
\label{eq:scaling law}
N=k_f\left(\frac{R_g}{a}\right)^{D_f}.
\end{equation}
Here $N$ is the number of constituents, each with radius $a$; $k_f$ is the fractal prefactor; $D_f$ is the fractal dimension, and $R_g$ is the radius of gyration defined by
\begin{eqnarray}
R_g^2&=&\frac{1}{N}\sum_{i=1}^N \left|\vec{r}_i-\vec{r}_0\right|^2,\\
\vec{r}_0&=&\frac{1}{N}\sum_{i=1}^N \vec{r}_i\,,
\end{eqnarray}
where $\vec{r}_i$ is the position of the $i$th constituent.
The value of the fractal dimension can in theory vary between the two extremes $D_f=1$ (a thin, straight chain of particles) and $D_f=3$ (a homogeneous sphere).
The aggregation method we employ ensures that with every particle that is added to the aggregate the scaling law (Eq.~\ref{eq:scaling law}) is fulfilled exactly \cite{Filippov}. For all aggregates we choose the fractal prefactor $k_f=2$. Different values of $k_f$ result in slightly different fractal aggregate shapes. By comparing the results obtained for various aggregates using different values of the fractal prefactor we find that the results are not very sensitive to the exact value chosen.

\begin{figure*}[!t]
\resizebox{\hsize}{!}{\hspace{5cm}\includegraphics{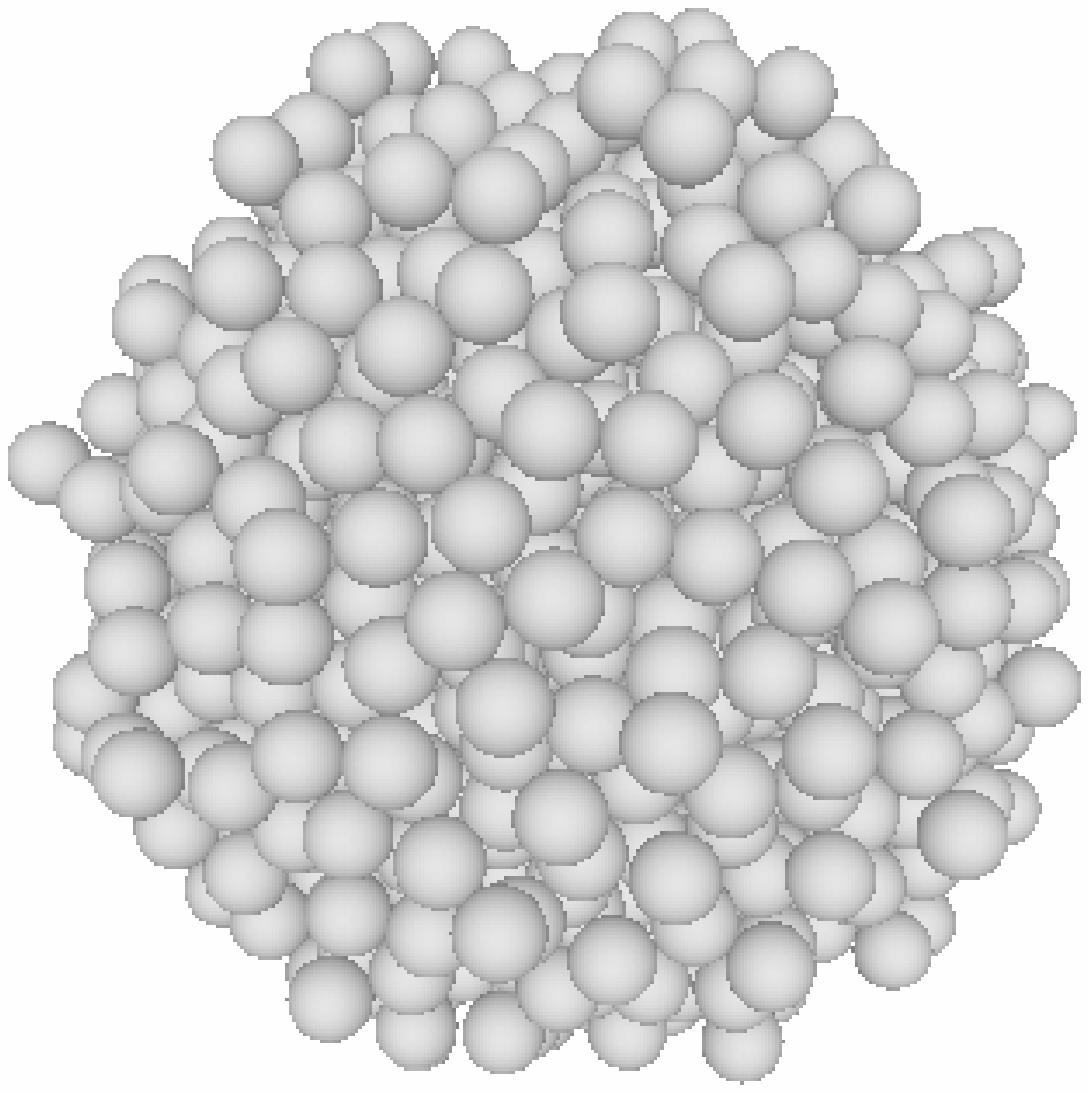}\includegraphics{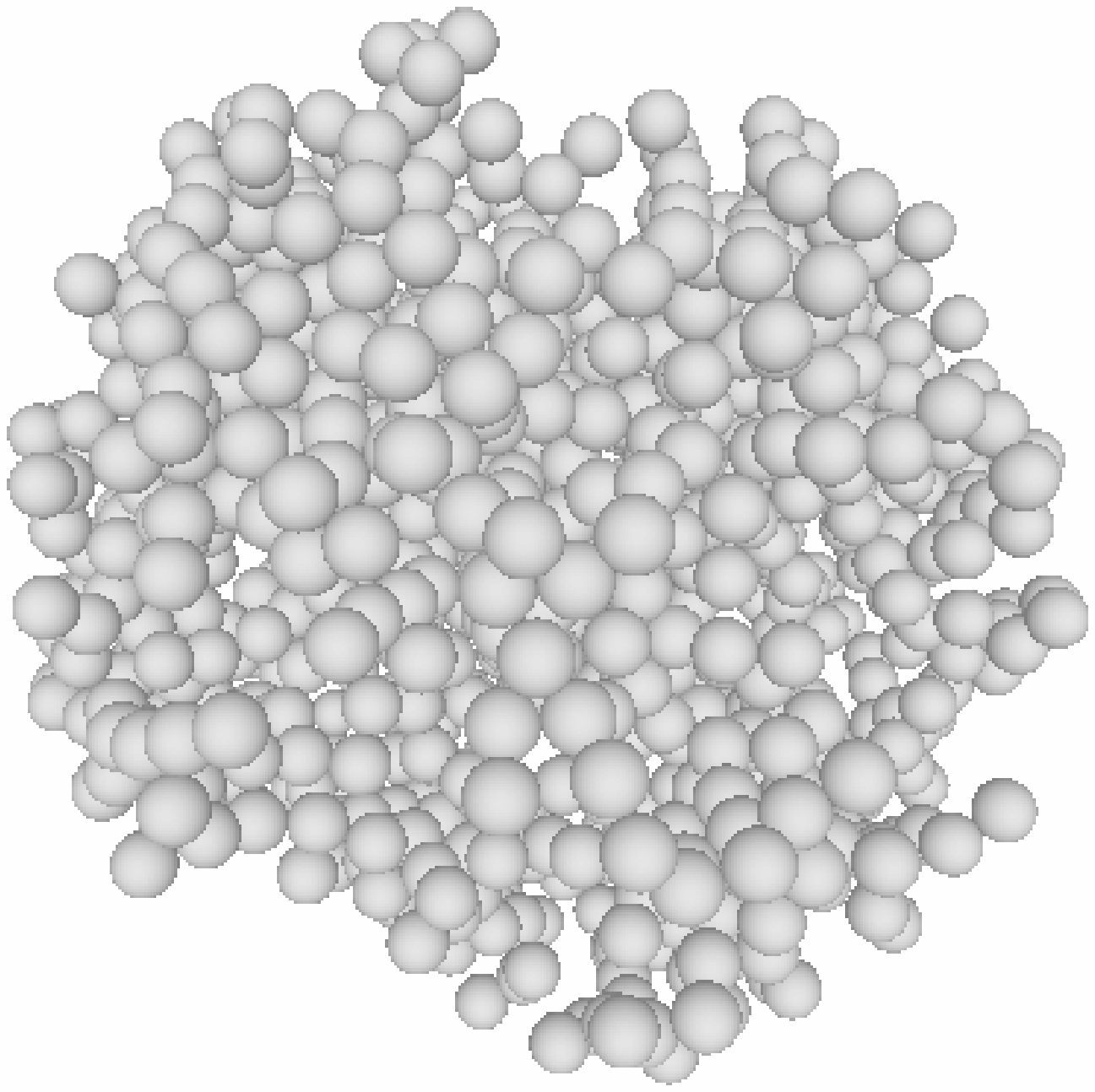}\includegraphics{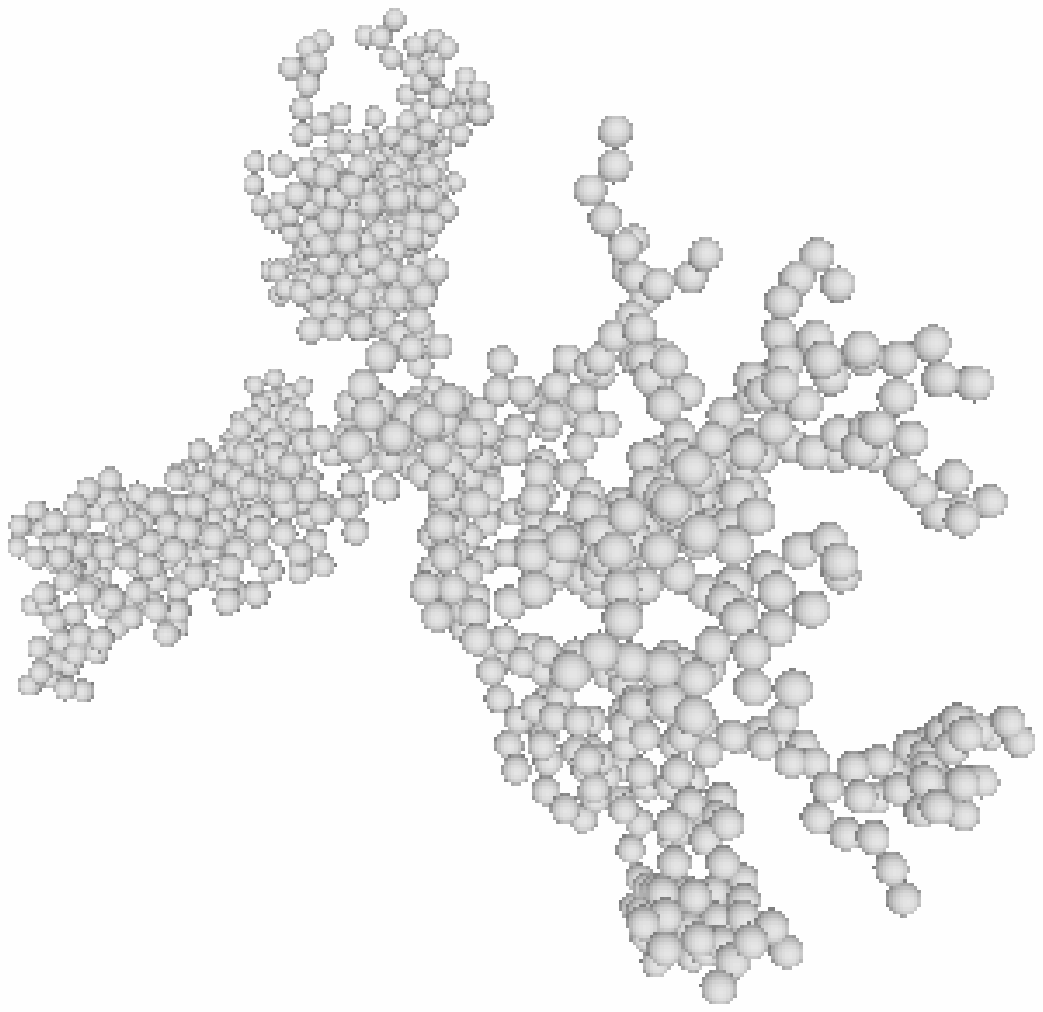}\includegraphics{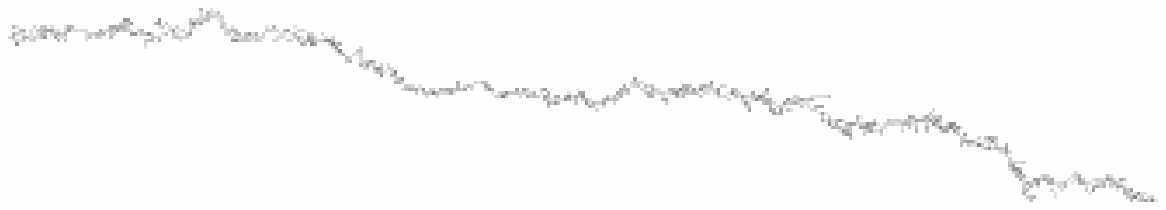}\hspace{1cm}}
\resizebox{\hsize}{!}{\includegraphics{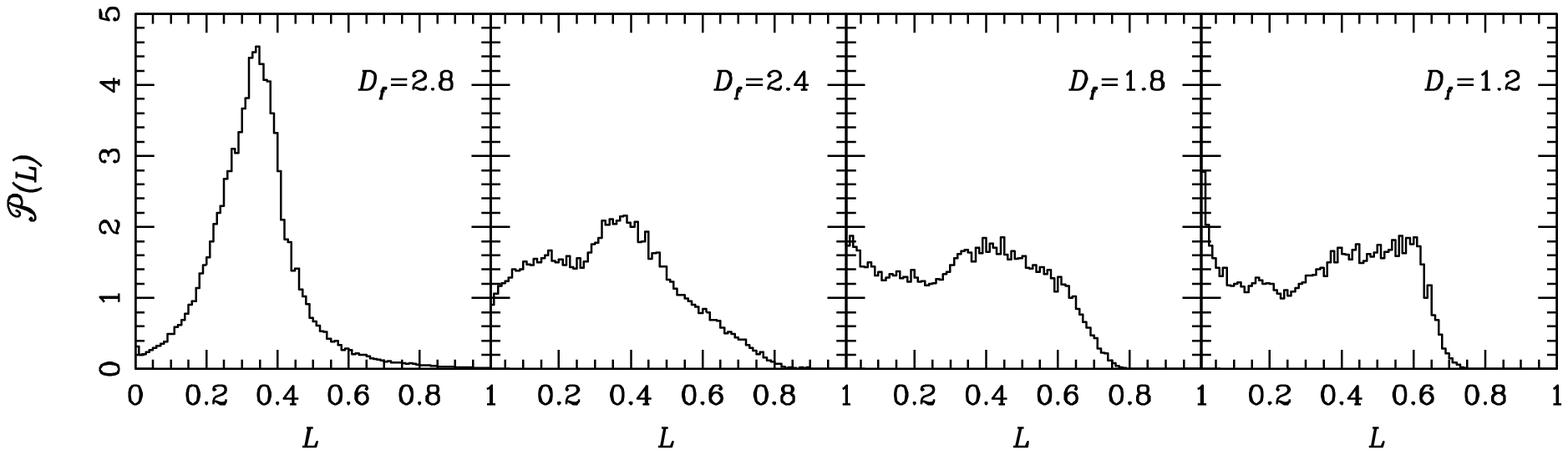}}
\resizebox{\hsize}{!}{\includegraphics{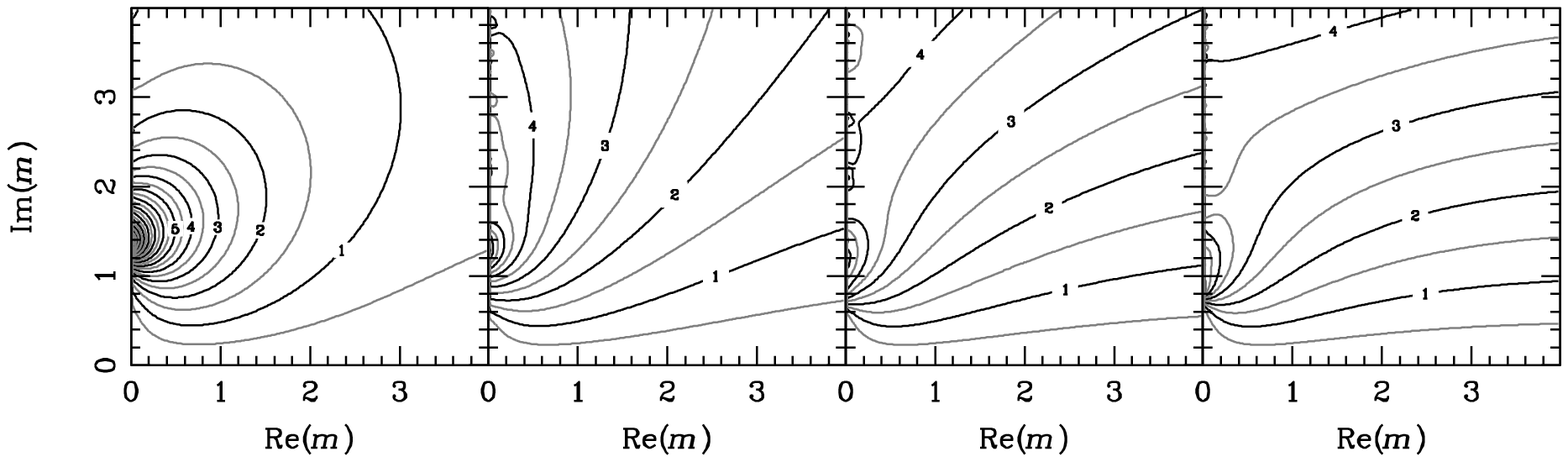}} 
\caption{Same as Fig.~\ref{fig:gaussian spheres} but for fractal aggregates.
From left to right the fractal dimension $D_f=2.8$, $2.4$, $1.8$ and $1.2$,
respectively. For all four fractal dimensions we considered an ensemble of $10$
particles with the same shape parameters but different seeds of the random
number generator. This results in $10$ fractal aggregates with the same fractal
dimension but different locations of the constituents. By averaging over these 
aggregates we reduce the effects of random variations of the particle shapes 
on the final results.} 
\label{fig:fractal aggregates} \end{figure*}

The structure of aggregates depends strongly on the precise mechanism by and environment in which these aggregates form.  The shape of aggregates formed in high density environments, for example small gold particles in aqueous suspension, is determined by the random-walk nature of the different particles upon approach.  This process can lead to aggregates with a fractal dimension of about 1.7 \cite{Weitz85}.  In low density environments like protoplanetary disks and molecular clouds in astrophysics, particles approach each other on straight lines before collisions.  If such clusters grow by addition of single grains to an aggregate (``Ballistic particle cluster aggregation'', BPCA), this process leads to aggregates with a fractal dimension of exactly 3 in the limit of large aggregates \cite{Ball84}, and slightly less than three for smaller aggregates. If on the other hand, the growth is approximately mono-disperse, i.e. each aggregate grows by collisions with aggregates of similar size (``Ballistic cluster cluster aggregation'', BCCA), the expected fractal dimension is between 1.8 and 2.1 \cite{kempf99}.  Almost linear aggregates ($D_f\approx 1$) may be formed from particles with electric or magnetic dipoles in external fields \cite{magnetic-I,magnetic-II}

Some typical examples of fractal aggregates are shown above the two panels of Fig.~\ref{fig:fractal aggregates}. The top panel of Fig.~\ref{fig:fractal aggregates} shows the ensemble and orientation averaged form-factor distributions of ensembles of fractal aggregates with different fractal dimensions. Each ensemble consists of fractal aggregates with the same fractal dimension and fractal prefactor, but with different values of the seed of the random number generator. Also shown in Fig.~\ref{fig:fractal aggregates} (bottom panel) is the average polarizability per unit volume as a function of the real and imaginary part of the refractive index of the particles.

It is apparent from Fig.~\ref{fig:fractal aggregates} that the imaginary part of the polarizability per unit volume of compact aggregates ($D_f=2.8$) resembles that of compact homogeneous spheres (see the left panel of Fig.~\ref{fig:sphere spheroid alpha}). When the fractal dimension is decreased, the resonance at $m=i\sqrt{2}$ disappears. Note the strong similarity between the bottom panels of Figs.~\ref{fig:gaussian spheres} and \ref{fig:fractal aggregates}. Although the Gaussian random spheres are very different from the fractal aggregates, the absorption properties for both classes of particles are quite similar.

\section{Discussion}
\label{sec:discussion}

\subsection{Comparison with distributions of simple shapes}

In order to model irregularly shaped particles, several distributions of simple shapes are presented in the literature. We consider here the widely used Continuous Distribution of Ellipsoids (CDE \cite{BohrenHuffman}) and the Distribution of Hollow Spheres (DHS \cite{2003A&A...404...35M}). For these distributions we compute the corresponding form-factor distributions and compare these to the form-factor distributions of irregularly shaped particles.

The CDE averages over all possible tri-axial ellipsoids with a specific weighting function (see \cite{2003A&A...404...35M}). The sum of the form-factors of a tri-axial ellipsoid averaged over all orientations has to be unity. Therefore, there are only two independent form-factors, $L_1$ and $L_2$ and one form-factor $1-L_1-L_2$. Since all form-factors have to be positive, another constraint is that $L_2<1-L_1$. The CDE averages over all possible ellipsoidal shapes by taking the distribution function ${\mathcal{P}(L_1,L_2)=2}$. The average polarizability then becomes
\begin{eqnarray}
\alpha&=&\frac{2}{3}\int_0^1 dL_1\int_0^{1-L_1}dL_2(\alpha(L_1)+\alpha(L_2)+\alpha(1-L_1-L_2))\nonumber \\
&=&2\int_0^1dL~(1-L)~\alpha(L).
\end{eqnarray}
Thus, the form-factor distribution according to the CDE is 
\begin{equation}
\label{eq:P(L) CDE}
\mathcal{P}(L)=2(1-L).
\end{equation}

In the DHS we average over the volume fraction, $f$, occupied by the central vacuum inclusion of a hollow sphere, while keeping the material volume of the particle constant. This implies that the outer radius of such a particle increases when $f$ increases. We consider the shape of a homogeneous hollow sphere to be different from that of a homogeneous sphere.
As can be verified from Eq.~(\ref{eq:Hollow sphere}) by substitution, the form-factor distribution of a single hollow sphere displays two features located at
\begin{eqnarray}
L_1&=&\frac{1}{2}-\frac{1}{6}\sqrt{8f+1}\,,\\
L_2&=&\frac{1}{2}+\frac{1}{6}\sqrt{8f+1}\,,
\end{eqnarray}
which have respective strengths
\begin{eqnarray}
w_1&=&\frac{1}{2}+\frac{1}{2\sqrt{8f+1}}\,,\\
w_2&=&\frac{1}{2}-\frac{1}{2\sqrt{8f+1}}\,.
\end{eqnarray}
From these equations it is already clear that the part $1/3<L<2/3$ in the form-factor distribution is not covered when using hollow spheres. From the above equations, the form-factor distribution for the DHS can be obtained and yields
\begin{equation}
\label{eq:P(L) DHS}
\mathcal{P}(L)=\left\{ \begin{array}{lrclcrcl}
\frac{3}{2}\left|2-3L\right|, \quad & 0\leq&L&\leq\frac{1}{3}\,\,\,&\mathrm{and}&\,\,\,\frac{2}{3}\leq&L&\leq1,\\
\\
0, & \frac{1}{3}<&L&<\frac{2}{3}\,.\\
\end{array}\right.
\end{equation}

\begin{figure*}[!t] 
\resizebox{\hsize}{!}{\includegraphics{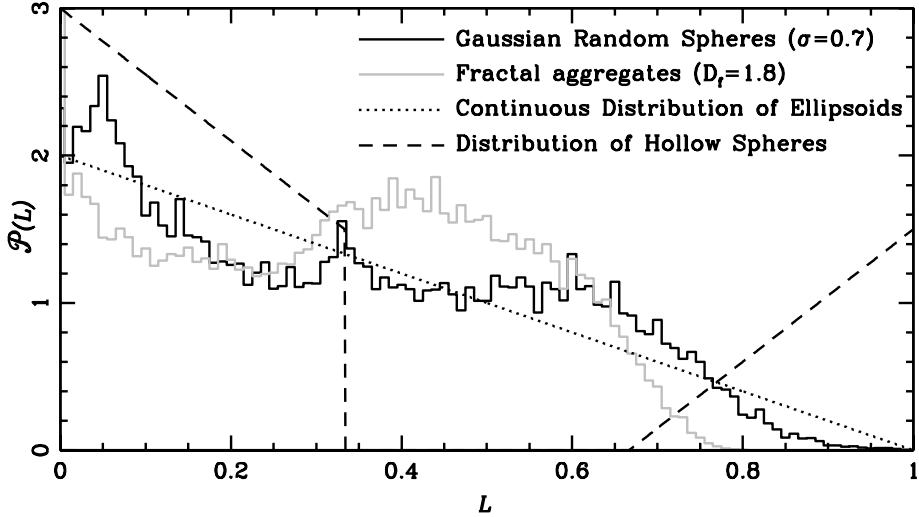}}
\caption{The form-factor distribution of various particle shape distributions.
The solid black line represents the distribution for an ensemble of randomly
oriented Gaussian Random Spheres with $\sigma=0.7$, the solid gray line
represents an ensemble of randomly oriented fractal aggregates with fractal
dimension $D_f=1.8$, the dotted line represents the Continuous Distribution of
Ellipsoids (CDE \cite{BohrenHuffman}) and the dashed line refers to the
Distribution of Hollow Spheres (DHS \cite{2003A&A...404...35M}). We chose
$\sigma=0.7$ for the Gaussian random spheres since more modest random spheres
($\sigma=0.1$ or $0.3$) still show too much resonance behavior (see
Fig.~\ref{fig:gaussian spheres}). The value of the fractal dimension $D_f=1.8$
is chosen because this is approximately expected from cluster-cluster
aggregation \cite{kempf99} and it represents a reasonably fluffy aggregate.
Note the remarkable similarity between the Gaussian Random Spheres with
$\sigma=0.7$ and the CDE form-factor distribution.} 
\label{fig:formfactors}
\end{figure*}

Fig.~\ref{fig:formfactors} shows the form-factor distributions of CDE (Eq.~\ref{eq:P(L) CDE}) and DHS (Eq.~\ref{eq:P(L) DHS}) together with those of an ensemble of Gaussian random spheres with $\sigma=0.7$ in random orientation and an ensemble of fractal aggregates with fractal dimension $1.8$ in random orientation. We can see that the form-factor distribution for the CDE and the Gaussian random spheres are in many respects very similar. This might explain the success of the CDE as a statistical representation of the absorption properties of irregularly shaped particles. We can also see from Fig.~\ref{fig:formfactors} that by using hollow spheres we do not cover the range of form-factors $1/3<L<2/3$ (cf. Eq.~\ref{eq:P(L) DHS}). Furthermore, the DHS shows a strong contribution of $L\gtrsim 0.8$ which is absent in the form-factor distributions of realistically shaped particles.

\subsection{Absorption spectra}

If we consider the wavelength dependence of the refractive index of the particle material, we can construct the absorption spectrum of the particle, i.e. its absorption cross section as a function of wavelength, using the form-factor distribution. In general, this absorption spectrum will display resonances, the exact positions and strengths of which are sensitive to the shape of the particles. As an example we consider the resonance of a single Lorentz oscillator. Its refractive index is given by
(see e.g. \cite{BohrenHuffman})
\begin{equation}
\label{eq:omega m}
m^2=m_0^2+\frac{f\omega_p^2}{\omega_0^2-\omega^2-i\gamma\omega}\,.
\end{equation}
In this equation $m_0>1$ is the real valued refractive index for $\omega\rightarrow\infty$, $f$ is the oscillator strength of the feature, $\omega_p$ is the plasma frequency, $\omega$ is the frequency of incident radiation and $\gamma$ is a damping factor. If $\omega$ is given in wavenumbers, we can express Eq.~(\ref{eq:omega m}) in wavelengths by using $\omega=\lambda^{-1}$. It can be shown that for a single form-factor, $L$, the absorption spectrum due to a single resonance given by Eq.~(\ref{eq:omega m}) displays a single feature at position $\omega_\mathrm{max}$ given by \cite{BohrenHuffman}
\begin{equation}
\label{eq:omega max}
\omega_\mathrm{max}^2=\omega_0^2+\frac{Lf\omega_p^2}{1+L(m_0^2-1)}\,,
\end{equation}
with a typical width $\gamma$. From this equation we can see that $\omega_\mathrm{max}$ is an increasing function of $L$. This means that low values of $L$ give a red feature (shifted towards long wavelengths, low frequencies), while high values of $L$ result in a blue feature (shifted towards short wavelengths, high frequencies). Eq.~(\ref{eq:omega max}) shows that the position of a resonance, $\omega_\mathrm{max}$, caused by a dust grain in the Rayleigh domain must obey the inequality
\begin{equation}
\omega_0^2\leq \omega_\mathrm{max}^2\leq \omega_0+\frac{f\omega_p^2}{1+(m_0^2-1)}\,.
\end{equation}

The maximum value of the imaginary part of the polarizability for a given Lorentz oscillator is also determined by the form-factor $L$. Furthermore,
\begin{equation}
\mathrm{max}\left[~\mathrm{Im}\left(\alpha\right)~\right]\propto \left(\frac{1}{1+L(m_0^2-1)}\right)^2.
\end{equation}
Thus, a low value of $L$ will result in a high value of the imaginary part of the polarizability. Combining this with the fact that low values of $L$ give a red feature, and that all of the form-factor distributions we computed give more weight to the lower values of $L$ than to the higher values, we conclude that, in general, particle shape effects tend to a broadening and a red-shift of the spectral features. This is in agreement with the findings presented in ref.~\cite{2003A&A...404...35M}.

\subsection{The statistical approach}

The method described in this paper provides a strong argument in favor of the main assumption of the statistical approach, namely that the optical properties of an ensemble of irregularly shaped particles can be represented in a statistical sense by the average properties of an ensemble of particles with the same composition, but with simple shapes. We have proved that for the absorption properties of particles in the Rayleigh domain, the statistical approach has an analytical basis given by the form-factor distribution. This implies that the average absorption cross section of an ensemble of arbitrarily shaped and arbitrarily oriented particles can be represented by a shape distribution of spheroidal particles with the same composition and in a fixed orientation.

We note that the form-factor distribution only provides the absorption cross sections, and not the scattering properties of small particles. It is in general not possible to find a distribution of spheroidal particles that gives both the absorption and the scattering cross section of an ensemble of arbitrarily shaped particles for every value of the refractive index. 
For an ensemble of ellipsoidal particles there is a relation between the scattering and the absorption cross section which is independent of the shape of the ellipsoids and is given by \cite{2003A&A...404...35M}
\begin{equation}
C_\mathrm{sca}=\frac{k^3V~|m^2-1|^2}{6\pi~\mathrm{Im}(m^2)}~C_\mathrm{abs}\,\,.
\end{equation}
From Eqs.~(\ref{eq:final Cabs}) and (\ref{eq:final Csca}) it is clear that for arbitrarily shaped particles such a simple relation which is independent of the particle shape can only be found in very specific cases.
This implies that in general there is no shape distribution of spheroids that can provide both the absorption and the scattering cross sections of an arbitrarily shaped particle for every value of the refractive index.

\subsection{Extrapolation to larger particles}
\label{sec:larger particles}

The form factor distribution can be used to construct for a given ensemble of irregularly shaped particles a shape distribution of simply shaped particles that has the same absorption properties in the Rayleigh domain. This distribution of simply shaped particles can be used for computations outside the Rayleigh domain. Combining the form-factor distribution of an irregularly shaped particle, $\mathcal{P}(L)$, with the equations for the form-factor of a spheroid with the field applied along the rotation axis, $L(a/b)$ (see Eq.~\ref{eq:L values}), the shape distribution of spheroids corresponding to this irregular particle shape is given by
\begin{equation}
\label{eq:Pab}
\mathcal{P}(a/b)=\mathcal{P}(L(a/b))~\left|\frac{dL(a/b)}{d(a/b)}\right|.
\end{equation}
Computations for spheroids in a fixed orientation can be computed relatively fast for large particles using for example the Separation of Variables Method (SVM \cite{1993Ap&SS.204...19V}) or the T-matrix method \cite{Mishchenko1996a}. Thus, Eq.~(\ref{eq:Pab}) provides a well founded choice for a distribution of spheroidal particles that can be employed for the computation of absorption properties of arbitrarily shaped particles with sizes outside the Rayleigh domain. However, it needs to be proven that this approach provides the correct optical properties in all cases.

\section{Conclusions}
\label{sec:Conclusions}

We have presented an easy to use method to compute the absorption and scattering properties of small particles with arbitrary shape, structure, orientation and composition. The method is based on a solution of the DDA equations in the Rayleigh domain. For a given geometrical shape of the particles, the solution has to be computed only once to obtain the absorption and scattering properties for arbitrary values of the refractive index. This provides a significant speedup of the computations in cases where calculations for many values of the refractive index have to be done. For example, it allows for fast computations of absorption spectra of arbitrarily shaped and arbitrarily oriented particles in the Rayleigh domain. Other practical applications are, for example, interpretation of atmospheric and radar measurements as well as calculations of the scattering matrix of small particles as a function of the scattering angle.

The method we use can be employed for calculating a form-factor distribution. This distribution uniquely determines the absorption properties of a particle or an ensemble of particles. Using the form-factor distribution, we have studied the basis of the statistical approach in the Rayleigh domain. Various shape distributions of particles can then exhibit the same absorption properties. Two ensembles of dust grains with the same composition and the same average form-factor distribution, but with different shape distributions have exactly the same absorption properties. Therefore, it is not possible to distinguish between these two ensembles from their absorption properties alone. Furthermore, for a given form-factor distribution it is trivial to obtain a distribution of spheroidal shapes with exactly the same average form-factor distribution. This provides a strong argument in favor of the fundamental assumption of the statistical approach, i.e. that the average absorption and scattering properties of an ensemble of irregularly shaped particles can be represented by an ensemble of simple shapes.

The form-factor distribution uniquely determines the absorption properties of arbitrarily shaped particles allowing for the construction of a shape distribution of spheroids with the same composition and a fixed orientation with the same average absorption cross section. It is, however, in general not possible to obtain a shape distribution of spheroidal particles with both the same average absorption and the same average scattering cross section. It needs to be investigated what the error on the computed scattering properties is when using the shape distribution of spheroids.

The conclusions presented in this paper are valid for particles in the Rayleigh domain. The form-factor distribution of an ensemble of irregularly shaped particles can be used to construct a shape distribution of spheroidal particles in a fixed orientation with exactly the same absorption properties in the Rayleigh domain. Since the absorption properties of spheroidal particles in a fixed orientation can be obtained relatively easy also for larger particles (i.e. outside the Rayleigh domain), this shape distribution provides a well founded choice for studying combined particle size and shape effects.

\subsection*{Acknowledgments}
We are grateful to M.~I. Mishchenko for valuable comments on an earlier version of this manuscript.

\end{document}